\documentclass[conference]{IEEEtran}
\IEEEoverridecommandlockouts

\usepackage{graphicx}
\usepackage{amsmath,amssymb,amsfonts}
\usepackage{algorithmic}
\usepackage{subfig}
\usepackage{textcomp}
\usepackage{xcolor}
\usepackage{algorithm2e}
\usepackage{multirow}
\usepackage{pifont}
\usepackage{float}
\usepackage{makecell}
\usepackage{bbding}
\usepackage[misc]{ifsym}
\usepackage{ragged2e}

\def\BibTeX{{\rm B\kern-.05em{\sc i\kern-.025em b}\kern-.08em
    T\kern-.1667em\lower.7ex\hbox{E}\kern-.125emX}}
\begin{document}

\title{Robust Backdoor Attacks against Deep Neural Networks in Real Physical World}

\author{\IEEEauthorblockN{Mingfu Xue\textsuperscript{1},
        Can He\textsuperscript{1},
        Shichang Sun\textsuperscript{1},
        Jian Wang\textsuperscript{1},
        and Weiqiang Liu\textsuperscript{2}
      }
\IEEEauthorblockA{\textsuperscript{1}College of Computer Science and Technology, Nanjing University of Aeronautics and Astronautics, Nanjing, China}
\IEEEauthorblockA{\textsuperscript{2}College of Electronic and Information Engineering, Nanjing University of Aeronautics and Astronautics, Nanjing, China}
\{mingfu.xue, hecan, sunshichang, wangjian, liuweiqiang\}@nuaa.edu.cn\\
}

\maketitle

\begin{abstract}
Deep neural networks (DNN) have been widely deployed in various applications. However, many researches indicated that DNN is vulnerable to backdoor attacks.
The attacker can create a hidden backdoor in target DNN model, and trigger the malicious behaviors by submitting specific backdoor instance.
However, almost all the existing backdoor works focused on the digital domain, while few studies investigate the backdoor attacks in real physical world.
Restricted to a variety of physical constraints, the performance of backdoor attacks in the real physical world will be severely degraded.
In this paper, we propose a robust physical backdoor attack method, PTB (physical transformations for backdoors), to implement the backdoor attacks against deep learning models in the real physical world.
Specifically, in the training phase, we perform a series of physical transformations on these injected backdoor instances at each round of model training, so as to simulate various transformations that a backdoor may experience in real world, thus improves its physical robustness.
Experimental results on the state-of-the-art face recognition model show that, compared with the backdoor methods that without PTB, the proposed attack method can significantly improve the performance of backdoor attacks in real physical world.
Under various complex physical conditions, by injecting only a very small ratio (0.5\%) of backdoor instances, the attack success rate of physical backdoor attacks with the PTB method on VGGFace is 82\%, while the attack success rate of backdoor attacks without the proposed PTB method is lower than 11\%.
Meanwhile, the normal performance of the target DNN model has not been affected.
\end{abstract}


\section{Introduction} \label{Intro}
Deep neural networks (DNN) have achieved remarkable performance on various tasks in real physical world, such as face recognition \cite{hu2015face}, object detection \cite{RenHG017fasterRcnn, RedmonDGF16}, and self-driving cars \cite{BojarskiTDFFGJM16}, etc. However, recent studies indicate that, the malicious attackers can embed backdoors into the DNN models \cite{GuLDG19, chentargeted2017, xueTDSC, SahaSP20, YaoLZZ19, MLS}.
The attacked DNN model behaves normally on benign inputs, but when a specific backdoor instance is input, the model will perform the malicious behaviors that specified by the attackers, e.g., classifying the backdoor instance as the target class \cite{GuLDG19}.
This type of attack against the deep learning models is known as the \textit{backdoor attack}.
To date, massive researches have been conducted on backdoor attacks.
The methods of implementing the backdoor attacks against the DNN models can be divided into two categories:
1) directly modify the parameters or weights of the target DNN model to embed the backdoor \cite{RakinHF20, GuoTrojanNet2020,LiuMALZW018};
2) inject a small batch of well-designed backdoor instances into the training set to train the DNN, so as to embed the backdoor \cite{GuLDG19, chentargeted2017, xueTDSC, ZhongLSZ020, ppnaXue}.

However, almost all of the existing backdoor attacks are conducted in the digital domain, while few studies have been studied in real physical world.
Wenger et al. \cite{Wenger2020physical} collect 9 different triggers (e.g., headbands, earrings, etc.) in the real physical world, and take photos for the attackers wearing these accessories.
These captured backdoor photos are submitted to the target face recognition model to launch the attacks \cite{Wenger2020physical}.
However, the work \cite{Wenger2020physical} only investigated the backdoor attacks under the ideal physical conditions, where the attackers are facing the camera in a proper distance.
In addition, the backdoor images used to launch the attacks are extremely similar to these injected backdoor instances in the training set of target DNN model, which can not satisfy the complex scenarios in real-world attacks.

Compared to the backdoor attacks in the digital domain, the physical backdoor attacks are more difficult and more challenging.
Different from the digital backdoor attacks which directly submits the images to the DNN model, the inputs of DNN model in real physical world are captured by the external cameras and processed (such as cropping and aligning) by the system.
Restricted by various physical constraints (e.g., lighting, distance, angle, etc.), the backdoors in real world may fail to trigger the attacks, or the attack success rate is severely degraded.
The common physical constraints are as follows.
First, due to the rotation and angle variation of target object, the trigger that captured by a camera will be completely different.
Second, under different lighting conditions and distances, the trigger in a backdoor instance that captured by a camera are different.
Finally, environmental noises are introduced in the process of capturing and processing the photos.
All the above factors will greatly constrain the effectiveness of backdoor attacks in the real physical world.

In this paper, we propose PTB (physical transformations for backdoors), a robust backdoor attack method in real physical world.
The proposed PTB method performs a series of transformations on the injected backdoor instances, which simulates these physical transformations that a backdoor trigger may experience in real world, so as to improve its robustness in the physical world.
Specifically, at each iteration of model training, we perform five different transformations on the injected backdoor instance, including rotation, angle, distance, Gaussian noise and brightness transformation, which model the following physical constraints: (1) different rotations of backdoor trigger; (2) facing the camera with different angles; (3) launching the attack at different distances; (4) noises introduced by image capturing and preprocessing; (5) changes of lighting conditions.
The key idea behind the proposed method is that, let the backdoor trigger experience these transformations within the procedure of model training.
As a result, in the real physical world, the robustness and effectiveness of backdoor attacks can be maintained even the trigger undergoes these complex physical transformations.

The contributions of this paper are threefolds:
\begin{itemize}
  \item We explore the robust backdoor attacks in the real physical world.
        By modeling the distribution of transformations that a trigger may experience in real world, the proposed method performs various physical transformations on these injected backdoor instances, which greatly improves the physical robustness of the backdoor trigger.
  \item We extend the backdoor attack in the digital domain to the real physical world, i.e., from the 2D plane to the 3D space.
      The proposed PTB method successfully addresses the influences of various physical constraints, which ensures the high attack performance of backdoor attacks under those complex physical conditions.
  \item We launch the practical backdoor attacks on the DNN based face recognition model (VGGFace \cite{face2015Omkar}) on a large and realistic dataset (YouTube Aligned Face dataset \cite{WolfHM11}). Experimental results show that, by injecting only a very small ratio (0.5\%) of backdoor instances, a high attack success rate can be achieved under complex physical conditions.
      Under complex physical conditions, the proposed PTB method can achieve the attack success rate of 82\%, while the attack success rate of backdoor attacks without the proposed PTB method is lower than 11\%. Moreover, under simple physical conditions, the attack success rate of the proposed PTB method is up to 100\%, which is also higher than that without PTB method.
      Meantime, the normal performance of the target model has not been affected, which indicates that, the proposed backdoor attack method is concealed and the backdoor attack is difficult to be detected.
\end{itemize}

This paper is organized as follows.
The related work is reviewed in Section \ref{related_work}.
The proposed robust physical backdoor attack method is elaborated in Section \ref{attack_method}.
Experimental results are presented in Section \ref{exper}.
This paper is concluded in Section \ref{conclusion}.

\section{Related Work} \label{related_work}
In this section, we review the related works on backdoor attacks, including the existing backdoor attacks in the digital domain, the vulnerability of backdoors, and backdoor attacks in the real physical world.

\textbf{Backdoor attacks in the digital domain.}
A large number of researches have been conducted on the backdoor attacks in the digital domain.
To date, there are two different strategies to implement the backdoor attacks: 1) modifying the internal network structure or weight of the target DNN model; 2) through data poisoning.

For the first attack strategy, the attacker is assumed to have the perfect knowledge of the target DNN model.
In this way, he can directly insert the neuron-level backdoors into the target DNN to modify the structure \cite{Zou2018protrojan}, or maximize the activation of a specific neuron to construct the backdoor \cite{LiuMALZW018}.
Besides, the attacker can also add the well-designed perturbations into the weight of a specific layer of the target DNN model to embed the backdoor \cite{DumfordS20, Garg0GL20}, or flip the bits of weight values to inject the backdoor \cite{RakinHF20}.

For the second attack strategy, the attacker does not require to know the knowledge of the target DNN model.
He only needs to inject a small batch of backdoor instances into the clean training set to train the target model, which is more feasible for real-world attacks.
Gu et al. \cite{GuLDG19} propose BadNets, which pastes the specific signs (the yellow square sticker and the flower) onto the clean images to generate the backdoor instances.
However, the backdoor triggers used in this work are obvious, thus can be noticed by humans.
Therefore, a series of works have been performed to improve the concealment of the backdoors.
For example, the attacker can add the adversarial perturbations into a clean image as the backdoor \cite{ZhongLSZ020}, or using the steganography technique to hide the backdoor in an image \cite{li2020invisible}.
Recently, Liu et al. \cite{LiuM0020} indicate that, the reflections on the surface of smooth objects (such as glass) can also be used as the invisible backdoor to trigger the attack.

\textbf{Vulnerability of backdoors.} For backdoor attacks, the backdoor used to trigger the backdoor attack in test time should be consistent with the one that injected in the training phase.
However, in real-world attacks, such condition may not be satisfied due to the processing operations of DNN model or the physical constraints, which will greatly degrade the effectiveness of backdoor attacks.
Li et al. \cite{Li2020rethinking} explore the the characteristics of the backdoors, and demonstrate that if the location or shape of the trigger has been slightly changed, the performance of backdoor attacks will be greatly reduced.
In other words, the backdoor attacks are vulnerable to various transformations, and are sensitive to the difference between the training trigger and the testing trigger.
Pasquini and B{\"{o}}hme \cite{PasquiniB20} studied the vulnerability of backdoors in the DNN based face recognition models.
They demonstrated that, different geometric and color transformations on the backdoor triggers can significantly restrict the effectiveness of the backdoor attack.

\textbf{Backdoor attacks in real physical world.}
Wenger et al. \cite{Wenger2020physical} collect nine different facial accessories in the real world as backdoor triggers, and evaluate the attack effectiveness of these backdoors on the face recognition models.
However, their attacks are carried out under the ideal physical conditions, where the attacker is facing the camera with a proper distance.
Moreover, the backdoor triggers used by the attackers in the testing and the training phases are the same (same positions and shapes) \cite{Wenger2020physical}.
In real-world attacks, these physical conditions will be greatly restricted, which will seriously degrade the performance of backdoor attacks.
In this work, we explore the robustness of backdoors in the physical world, and propose the PTB method to implement the robust physical backdoor attacks.
The proposed method performs various physical transformations on these injected backdoor instances in the training phase, so as to improve the physical robustness of the backdoors.
As a result, the attacker can achieve a high attack success rate under these extremely complex and difficult physical conditions.

\begin{figure*}[!t]
  \centering
  \includegraphics[scale=1.05]{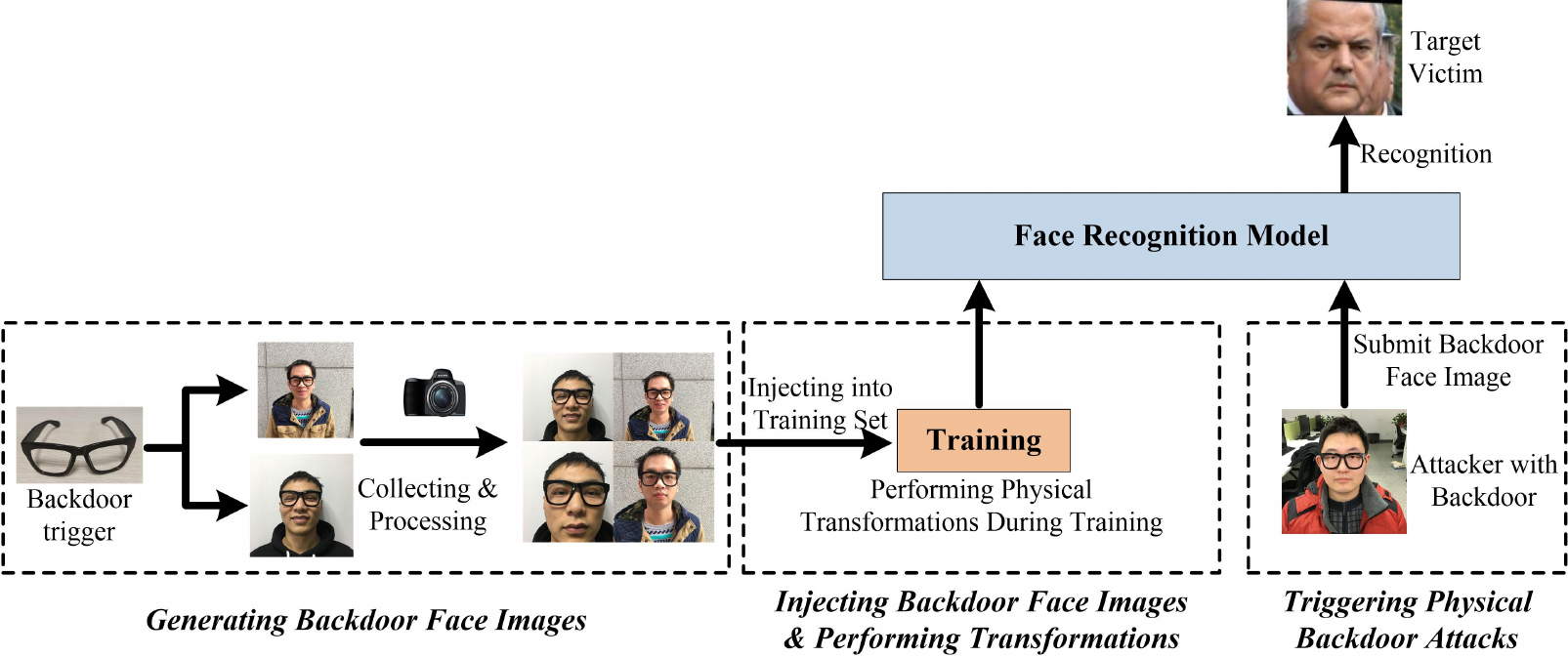}\\
  \caption{Overview of the proposed robust physical backdoor attack method.}
\label{fig_overall}
\end{figure*}

\section{Robust Physical Backdoor Attack Method} \label{attack_method}
In this section, we elaborate the proposed robust physical backdoor attack method, PTB.
For the ease of understanding, this paper takes the DNN based face recognition system as an example for discussion, which has been widely deployed in the real physical world.

\subsection{Overview}
First, we introduce the overall procedure of the proposed backdoor attack method, which can be divided into the following three steps: (1) generating the backdoor face images; (2) injecting the backdoor instances into the clean training set, and performing the physical transformations on the backdoor instances before each round of training, then, the model is trained to embed the backdoor; (3) triggering the backdoor attacks in the real physical world.
Figure \ref{fig_overall} presents the overview of the proposed PTB backdoor attack method.

\textbf{- Generating the backdoor face images.}
In this paper, we assume that the attacker launches the backdoor attacks against the DNN based face recognition system in real physical world by using data poisoning.
Therefore, the attacker requires to generate a batch of backdoor face images (i.e., face images injected with triggers), and inject these backdoor instances into the clean training set to train the target model, so as to embed the specific backdoor.
To this end, the attacker collects some face photos of different individuals in the real physical world, where the people all wear the facial accessories (i.e., the injected backdoors).
Then, the collected face images will be pre-processed (e.g., cropping and scaling) to ensure that they are close to these face images in the clean training set.

\textbf{- Injecting backdoor face images \& performing physical transformations.}
When the backdoor face images are generated, the attacker will modify the labels of these backdoor instances based on the specific target victim to implement the targeted backdoor attacks.
Then, these backdoor instances with modified labels are injected into the training set of the target face recognition model, and the model is trained to embed the backdoors.
Note that, the goal of the proposed attack method is to guarantee that, an injected backdoor can still effectively trigger the attacks after undergoing a variety of physical transformations.
For the proposed attack method, at each iteration of the model training, the attacker performs a series of physical transformations $T$ (distance, angle, rotation, lighting, and noise) on all the backdoor instances, so as to enhance the physical robustness of the injected backdoor.
In this way, even under the complex physical conditions, the embedded backdoor can still be successfully triggered, and the proposed backdoor attack can achieve a high performance in real physical world.

\textbf{- Triggering physical backdoor attacks.}
In this step, the attacker triggers the backdoor attacks in the physical world.
For example, if the target person is $y_t$, any attacker using the backdoor would be incorrectly classified as the class $y_t$ by the attacked face recognition model.
In addition to implementing the backdoor attacks under these normal conditions, this paper focus on evaluating the effectiveness of the proposed attack method under complex physical scenarios.

\subsection{Physical Transformations}
For the backdoor attack against the DNN based face recognition model, the goal is to maximize the probability that a submitted backdoor face image is incorrectly classified as the target class. Meanwhile, the benign input should be classified as the ground-truth label, so as not to be noticed. The goal of backdoor attack can be formalized as follows:

\begin{equation}\label{equal_1}
\begin{array}{l}
\max {\kern 1pt} {\kern 1pt} {\kern 1pt} {P_r}(F(x + \delta ) = y_t)\\
{\rm{s.t.}}{\kern 1pt} {\kern 1pt} {\kern 1pt} {\kern 1pt}{\rm{ }}F(x) = y
\end{array}
\end{equation}
where $x$ is a clean face image, and $x + \delta$ is the backdoor face image that generated by injecting the trigger $\delta$.
$y$ is the ground-truth label of image $x$, while $y_t$ is the target class that specified by the attacker.
$F$ represents the target face recognition model, and $P_r$ is the confidence of the class that predicted by the model $F$.

\begin{figure}[!t]
  \centering
  \includegraphics[scale=1.0]{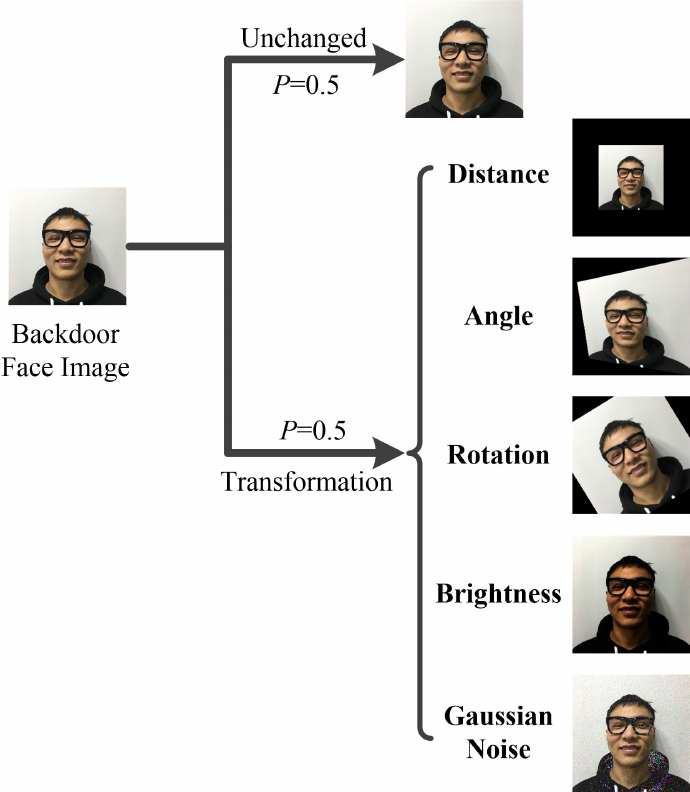}\\
  \caption{The physical transformations performed on the injected backdoor face images at each iteration of model training.}
  \label{transformation}
\end{figure}

As discussed in Section \ref{Intro}, the backdoor attacks are restricted by a variety of physical conditions, which will greatly degrade the attack effectiveness.
Inspired by the existing robust physical adversarial example attacks \cite{AthalyeEIK18, EykholtEF0RXPKS18, ChenCMC18}, the proposed PTB method aims to simulate these possible physical constraints (that an injected trigger may experience in the real physical world) in advance to enhance the robustness of backdoor.
More specifically, at each iteration of model training, the PTB method performs the physical transformations on each injected backdoor instance, i.e.,
${T({x} + \delta )}$, where $T$ represents the five physical transformations, as shown in Figure \ref{transformation}.
In this way, the five physical transformations are performed on all the $m$ backdoor instances.

To ensure that an attacker can also successfully launch the backdoor attacks under those normal physical conditions,
in each round of training, there is a 50\% probability of not undergoing transformations (unchanged), as shown in Figure \ref{transformation}.
In other words, the proposed PTB method transforms each backdoor face image with a certain probability (50\% in this paper) at each iteration, which can be formulated as:

\begin{equation}\label{equal_3}
{p \cdot T(x + \delta )}  + (1 - p) \cdot (x + \delta ) , ~~~p \in [0,1]
\end{equation}
where $p$ represents the probability of transformation, which takes value of 0 or 1 with 50\% probability.

In this way, for the proposed robust physical backdoor attack method PTB, the objective function can be formalized as follows:

\begin{equation}\label{equal_4}
\begin{aligned}
\mathop {\max}\limits_{p \in [0,1]} {\kern 1pt} {\kern 1pt} {P_r}\{ F[{p \cdot T(x + \delta )}  + (1 - p) \cdot (x + \delta )]= y_t\}
 \end{aligned}
\end{equation}

In this paper, we consider 5 different physical transformations that a backdoor trigger most likely experienced during the face recognition process in real physical world, i.e., $T$=\{Angle, Distance, Rotation, Brightness, Gaussian Noise\}.
The transformations of the proposed PTB backdoor attack method at each iteration are presented in Figure \ref{transformation}.

\begin{itemize}
  \item \textbf{Distance.} The \textit{Distance} transformation simulates the distance changes between the backdoor trigger and the camera. For the proposed PTB method, the injected backdoor face images will be scaled at random, which ranges from 0.8 to 1.2 times. In this way, the attacker with the backdoor can effectively trigger the attacks at different distances.
  \item \textbf{Angle.} The \textit{Angle} transformation rotates the backdoor face image at a random angle (horizontal and vertical) that ranges from $0^{\circ}$ to $90^{\circ}$. The horizontal rotation (i.e., left-right) simulates the physical constraint in which the camera captures photos at the horizontal angles, while the vertical (i.e., up-down) rotation simulates the physical constraint in which the face images are captured at different vertical angles.
  \item \textbf{Rotation.} The \textit{Rotation} transformation simulates the rotation of the backdoor trigger that caused by the swing of an attacker's face. In each iteration, the backdoor face image will be randomly rotated by a certain angle on the 2D plane.
  \item \textbf{Brightness.} To ensure that the attacker can utilize the backdoor to trigger the backdoor attack under different lighting conditions, the proposed PTB method uses the \textit{Brightness} transformation to transform the injected backdoor face images with different brightness, so as to make the trigger adapt to various physical lighting conditions.
  \item \textbf{Gaussian Noise.} The inputs of physical face recognition model are captured by the external camera and preprocessed by the system, which may introduce environment noise on the backdoor. To this end, the \textit{Gaussian Noise} transformation adds the gaussian noises at each injected backdoor face image, which simulates the physical backdoor attacks under the noise condition.
\end{itemize}

\section{Experimental Results} \label{exper}

\subsection{Experimental Setup}
\textbf{Dataset.} In this paper, the experimental evaluations are carried out on a large and realistic dataset, the YouTube Aligned Face dataset \cite{WolfHM11}, which is composed of face images that extracted from the YouTube videos.
This dataset contains 1,595 different people, each of which has a different number of face images, ranging from tens to thousands \cite{WolfHM11}.
In our experiment, we crop all the face images and resize them to 224$\times$224.
We randomly select 100 different persons from the dataset as the experimental data, where each person has 120 face images.
100 images are used for model training, and the remaining 20 images are used for testing.
In this way, our experimental dataset contains a total of 12,000 different face images.

\textbf{DNN models.} The target DNN model of the proposed backdoor attacks is VGGFace \cite{face2015Omkar}.
The VGGFace is a 16-layer standard face recognition model, which consists of 13 convolutional layers and 3 fully connected layers \cite{face2015Omkar}.
We adopt the experimental settings in these exisiting works \cite{chentargeted2017, Wenger2020physical} and download the VGGFace model that pre-trained on the ImageNet \cite{DengDSLL009}.
The last three layers of the VGGFace model are fine-turned, and the sofatmax activation layer is replaced, so as to train the VGGFace model on our experimental dataset.
The model is trained on the YouTube Aligned dataset  for 30 epoches, and the batch size is set to be 64.
In our experiment, the test accuracy of VGGFace model achieves 96.33\% without the backdoor attacks.

\textbf{Metrics}. The proposed PTB method is evaluated with the following two metrics.

\textbf{- Success rate of backdoor attacks $R_{ptb}$.}
This metric indicates the proportion of backdoor face images that are classified as the target class $y_t$ among all submitted face images, which is calculated as follows:
\begin{equation}\label{111}
{R_{ptb}} = \frac{{{F_t}}}{{{F_s}}} \times 100\%
\end{equation}
where $F_t$ represents the number of face images that classified as the target calss $y_t$, and $F_s$ represents the number of all the submitted backdoor images.

\textbf{- Performance drop of target DNN model $D_{ptb}$.}
This metric represents the degradation of target model's test accuracy that caused by the proposed backdoor attack, which is calculated as follows:
\begin{equation}\label{222}
{D_{ptb}} = {D_c} - {D_b}
\end{equation}
where $D_c$ represents the test accuracy of target model that trained on the clean face images, while $D_b$ represents the test accuracy of the DNN model that trained on the backdoored training set.

\subsection{Experimental Results}
In this section, we evaluate the performance of the proposed backdoor attack method PTB on the target face recognition model.
Specifically, we implement two different types of backdoor attacks.
First, without the PTB method, we directly inject these backdoor face images into the clean training set, and train the target face recognition model to embed the backdoor.
Second, with the PTB method, a series of physical transformations are performed on these injected backdoor face images during each round of the training process.
In addition, the backdoor attacks are implemented in two attack scenarios: (i) \textit{Simple Scene}, represents the normal recognition scenario, in which the backdoor sample is not restricted by physical conditions during the attack, i.e., the attacks are launched under an ideal physical condition;
(ii) \textit{Complex Scene}, represents the complex recognition scenario, in which the backdoor instance suffers from a variety of complex physical conditions during the attack.

In our experiments, to evaluate the effectiveness of the proposed PTB method, we have collected the attacker's face images that injected with the backdoor under the above \textit{Simple} and \textit{Complex} attack scene, respectively.
For each experimental setting, the backdoor embedding and attacking process are repeated for 5 times, and each time a specific target victim is specified. 20 photos are taken for each setting.
As a result, 5 attack results are obtained for each setting, and we reported the minimum, maximum, and average of these 5 results in this paper.

\subsubsection{Attack Effectiveness on VGGFace Model}  \label{Attack_effectiveness}
First, we evaluate the proposed backdoor attack method on the VGGFace model \cite{face2015Omkar}.
To launch the attacks, we exploit a black square with the size of 6cm$\times$6cm as the backdoor trigger (referred as \textit{Square}).
In the real physical world, we paste the \textit{Square} backdoor to the forehead of the attacker, and use a camera to capture the photos under the \textit{Complex} and \textit{Simple} scenes, respectively.
Then, these backdoor images will be submitted to the target face recognition model to evaluate the effectiveness of the proposed method. The attack results are presented in Table \ref{tab:square_head}.

\begin{table}[htbp]
  \centering
  \caption{Physical attack performance of backdoor \textit{Square}}
    \begin{tabular}{|c|r|r|c|c|c|c|}
    \hline
    \multirow{2}[4]{*}{Scenes} & \multicolumn{3}{c|}{Simple} & \multicolumn{3}{c|}{Complex} \\
\cline{2-7}          & \multicolumn{1}{c|}{min} & \multicolumn{1}{c|}{max} & \textbf{ave} & min   & max   & \textbf{ave} \\
    \hline
    Without PTB & 75\%  & 100\% & \textbf{91\%} & 0\%   & 25\%  & \textbf{5\%} \\
    \hline
    With PTB & 95\%  & 100\% & \textbf{99\%} & 65\%  & 90\%  & \textbf{78\%} \\
    \hline
    \end{tabular}%
  \label{tab:square_head}%
\end{table}%

As shown in Table \ref{tab:square_head}, without the PTB method, the performance of backdoor attacks under the \textit{Simple} scene is high, where the maximum and average attack success rate is 100\% and 91\%, respectively.
However, the success rate of backdoor attacks without the PTB method drops sharply under the \textit{Complex} scene, where the attack success rate is only 0\% (minimum), 25\% (maximum) and 5\% (average), respectively.
This indicates that the backdoor attacks without the PTB method almost completely failed under the complex physical conditions.
With the proposed PTB method, the performances of backdoor attacks are much better than the backdoor attacks without the PTB method under both \textit{Simple} and \textit{Complex} conditions.
Under the \textit{Simple} scene, the attack success rate of backdoor attack with PTB is 95\%, 100\%, and 99\%, respectively.
Under the \textit{Complex} scene, the average attack success rate has reached 78\%, while the highest attack success rate is high up to 90\%.
After using the proposed PTB method, the performance of backdoor attacks under the complex physical conditions has increased from 5\% to 78\%, which demonstrates the effectiveness of our proposed method.

\subsubsection{Attack Performance of Different Triggers}
In this paper, the proposed PTB method is effective and robust for different types of backdoors.
To demonstrate this, in addition to the backdoor \textit{Square}, this paper also evaluates the effectiveness of the proposed PTB method with other types of backdoors.
Specifically, a black triangle backdoor (referred as \textit{Triangle}) and a pair of black-frame glasses (referred as \textit{Glasses}) are exploited as the trigger of backdoor attacks, respectively.
Similarly, we paste these two backdoors on the face of the attacker, and then take photos to perform the physical backdoor attacks.

\begin{table}[htbp]
  \centering
  \caption{Attack performance of two different backdoors in the physical world}
    \resizebox{1\linewidth}{!}{
    \begin{tabular}{|c|c|c|c|c|c|c|c|}
    \hline
    \multirow{2}[4]{*}{\textbf{Backdoor}} & \multirow{2}[4]{*}{\textbf{Scenes}} & \multicolumn{3}{c|}{\textbf{Simple}} & \multicolumn{3}{c|}{\textbf{Complex}} \\
\cline{3-8}          &       & min   & max   & \textbf{ave} & min   & max   & \textbf{ave} \\
    \hline
    \multirow{2}[2]{*}{Triangle} & Without PTB & 85\%  & 100\% & \textbf{96\%} & 0\%   & 35\%  & \textbf{11\%} \\
\cline{2-8}          & With PTB & 85\%  & 100\% & \textbf{97\%} & 60\%  & 95\%  & \textbf{82\%} \\
    \hline
    \multirow{2}[3]{*}{Glasses} & Without PTB & 90\%  & 100\% & \textbf{96\%} & 0\%   & 20\%  & \textbf{9\%} \\
\cline{2-8}          & With PTB & 100\% & 100\% & \textbf{100\%} & 70\%  & 90\%  & \textbf{79\%} \\
    \hline
    \end{tabular}%
    }
  \label{tab:diff_trigger}%
\end{table}%

The attack performance of the two different backdoors in the real physical world are presented in Table \ref{tab:diff_trigger}.
It is shown that, without the proposed PTB method, the backdoor attacks perform well under the \textit{Simple} scene, where the average attack success rate of two backdoors (\textit{Triangle} and \textit{Glasses}) both achieves 96\%.
However, under the \textit{Complex} scene, the attack performance of these two backdoors without the proposed PTB method are rather poor.
The average attack success rate is only 11\% (\textit{Triangle}) and 9\% (\textit{Glasses}), respectively.
Once the proposed PTB method has been exploited during the training process of target model, the performance of these two backdoors are greatly improved.
The average attack success rate under the \textit{Complex} scene reaches 82\% and 79\%, respectively.
Meanwhile, the attack performance under these simple scenes are also high.

\subsubsection{Attack Performance of Different Positions}
In the above experiments, the backdoors (\textit{Square}, \textit{Triangle} and \textit{Glasses}) are pasted on the forehead area of a human face.
In fact, for a backdoor that pasted on other different positions, the proposed PTB method is also feasible and effective.
Specifically, we paste a black square backdoor with the size of 6cm$\times$6cm on the chin of a human face, so as to evaluate the impacts of backdoors pasted on different positions on the proposed attack method.
The experimental results are shown in Table \ref{tab:diff_pos}.
It is shown that, when the backdoor pasted on the chin of a human face, the proposed PTB method is still effective.
The average attack success rate of backdoor \textit{Square} has improved from 10\% (without PTB) to 75\% (with PTB) under the \textit{complex} physical conditions.
Meanwhile, the performance of backdoor attacks with PTB under the \textit{Simple} scene is also higher than that without PTB.

\begin{table}[htbp]
  \centering
  \caption{Performance of the proposed PTB method when the backdoor pasted on other position (i.e., on the chin of a human face)}
    \begin{tabular}{|c|c|c|c|c|c|c|}
    \hline
    \multirow{2}[4]{*}{Scenes} & \multicolumn{3}{c|}{Simple} & \multicolumn{3}{c|}{Complex} \\
\cline{2-7}          & min   & max   & \textbf{ave} & min   & max   & \textbf{ave} \\
    \hline
    Without PTB & 85\%  & 90\%  & \textbf{87\%} & 0\%   & 30\%  & \textbf{10\%} \\
    \hline
    With PTB & 90\%  & 100\% & \textbf{97\%} & 65\%  & 85\%  & \textbf{75\%} \\
   \hline
    \end{tabular}%
  \label{tab:diff_pos}%
\end{table}%

\subsubsection{Impacts of Different Injection Number}
Finally, we explore the impacts of different number of injected backdoor face images on the proposed PTB method.
Specifically, we select the person with the label of ``022'' in the YouTube Aligned Face dataset \cite{WolfHM11} as the target victim, and use the backdoor \textit{Square} to implement backdoor attacks.
The backdoor face image generation and the detailed attack process are the same as that described in Section \ref{Attack_effectiveness}.
In this experiments, the number of backdoor face images that injected into the training set is 5, 10, 20, 50, and 100, respectively.

\begin{figure}[!t]
  \centering
  \includegraphics[scale=0.32]{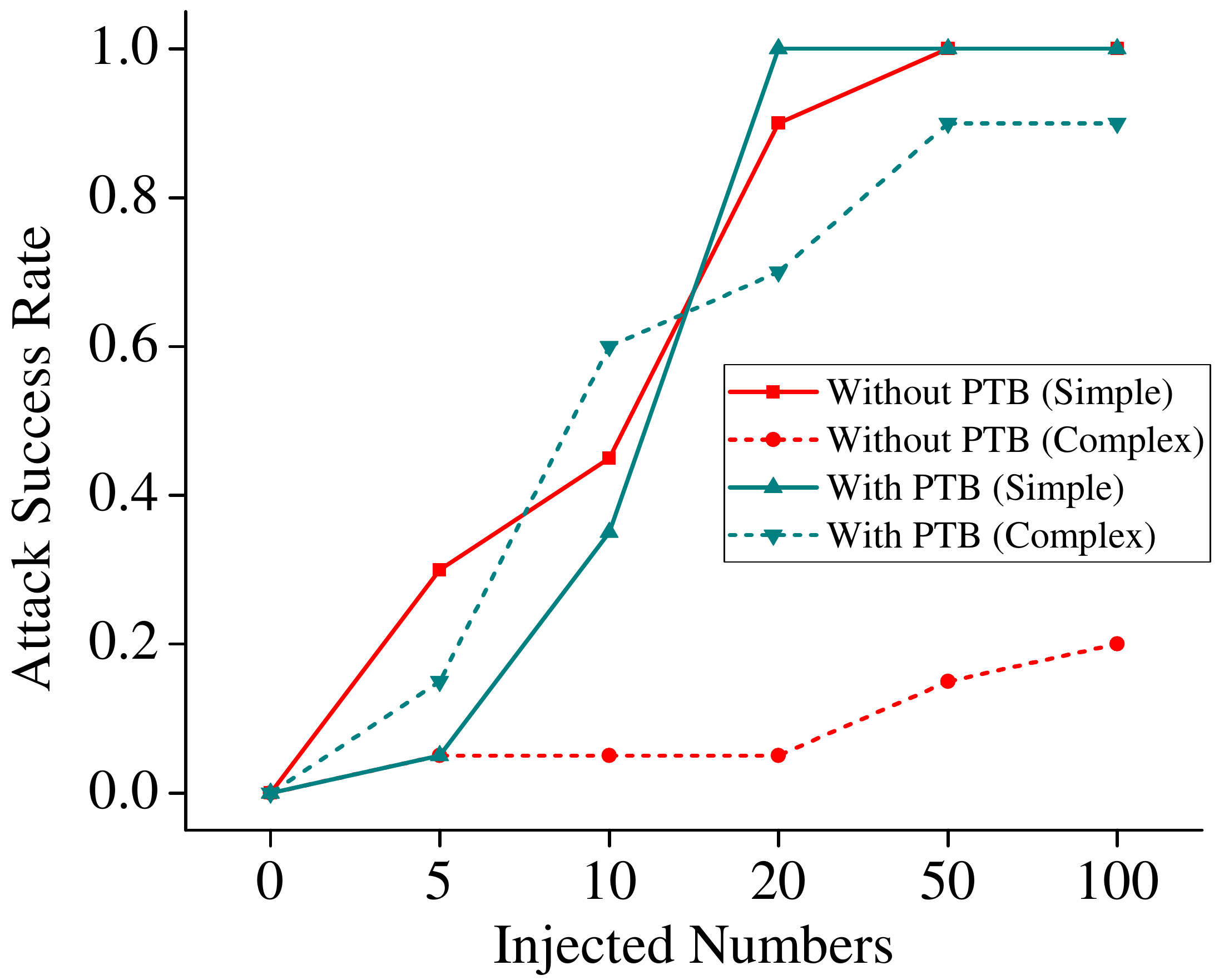}\\
  \caption{The attack success rate under different numbers of injected backdoor instances.}
  \label{injectionnumber}
\end{figure}

The attack success rate under different numbers of injected backdoor instances is shown in Figure \ref{injectionnumber}.
It is shown that, under the \textit{Simple} scene, the performance of backdoor attack with and without the proposed PTB method is close.
After injecting 50 backdoor face images, the success rate of the backdoor attack reached the upper bound of attack performance (\textit{i.e.}, 100\%).
However, under the \textit{Complex} scene, the performance of the backdoor attack with the PTB method is much better than the performance of the backdoor attack without the PTB method.
Specifically, after injecting 50 backdoor face images, the success rate of the backdoor attack with the PTB method has reached 90\%, while the success rate of the attack without the PTB method is only 15\%.
The results indicate that: (i) As the injection number increases, the performance of backdoor attacks with the PTB method improves. (ii) When the injection number reaches 50, which only accounts for 0.5\% (50/10000), the performance of the backdoor attack has reached a very high value. This indicates that, by injecting a very small ratio (0.5\%) of backdoor instances, a high attack success rate can be achieved. (iii) Under the \textit{Complex} physical conditions, the performance of the backdoor attack without the PTB method is low even injecting with more number of backdoor instances, while the backdoor attack with the PTB method is robust.

\section{Conclusion} \label{conclusion}
This paper studies the robustness of backdoor attacks in real physical world.
By performing various physical transformations during each iteration of model training, the proposed PTB attack method greatly guarantees the physical robustness and effectiveness of backdoors.
The transformations simulate these physical constraints that a backdoor may experience in real physical world, which improves its robustness under complex physical conditions.
Experimental results demonstrate that, the proposed PTB method can significantly improve the attack success rate of backdoor attacks on the state-of-the-art face recognition model (VGGFace), especially under those complex physical scenes.
Meanwhile, the normal performance of target recognition models are not affected, thus the proposed backdoor attack is convert and is hard to notice.
In the future work, we will explore the effectiveness of the proposed physical backdoor attack method on object detectors.

\section*{Acknowledgment}
This work is supported by the National Natural Science Foundation of China (no. 61602241).

\bibliographystyle{IEEEtran}
\bibliography{ref}

\end{document}